\documentclass[twocolumn,showpacs,prl]{revtex4}
\usepackage{amsmath}
\usepackage{graphicx}
\usepackage{dcolumn}
\usepackage{bm}

\begin{document}

\title{Spin-independent effective mass in a valley-degenerate electron system}
\author{Suhas Gangadharaiah and Dmitrii L. Maslov}
\date{\today}

\begin{abstract}
In a generic spin-polarized Fermi liquid, the masses of spin-up
and spin-down electrons are expected to be different and to depend
on the degree of polarization. This expectation is not confirmed
by the experiments on two-dimensional heterostructures. We
consider a model of an $N$-fold degenerate electron gas. It is
shown that in the large-$N$ limit, the mass is enhanced via a
polaronic mechanism of emission/absorption of virtual plasmons. As
plasmons are classical collective excitations, the resulting mass
does not depend on $N$, and thus on polarization, to the leading
order in $1/N$.  We evaluate the $1/N$ corrections and show that
they are small even for $N=2$.
\end{abstract}


\affiliation{ Department of Physics, University of Florida, P. O.
Box 118440, Gainesville, FL 32611-8440 }

\maketitle The observation of an apparent metal-insulator
transition in high-mobility Si
metal-oxide-semiconductor-field-effect-transistors (MOSFET's)
\cite{krav1} challenged the scaling theory of localization
\cite{abrahams}, which predicts that a two-dimensional (2D) system
undergoes only a continuous crossover between weak and strong
localization regimes. Although there has been a substantial
progress in understanding of transport and thermodynamic
properties of MOSFET's and other heterostructures
\cite{abrahams_rmp,pudalov_review}, the origin of the observed
phenomena is still a subject of discussion. Although a
conventional (dirty) Fermi-liquid (FL) theory \cite{zna,finn} can
account for many observed effects at least qualitatively and, in
some cases, quantitatively, there is also a number of non-FL
scenarios for the anomalous metallic state
\cite{chakravarty,spivak}. On the experimental side, the main
argument for the FL-nature of the metallic state is the
observation of quite conventional Shubnikov-de Haas (ShdH)
oscillations \cite{abrahams_rmp,pudalov_review}, which implies an
existence of well-defined quasiparticles albeit with the
renormalized effective mass $m^*$ and spin susceptibility
$\chi_s^*$.  The ShdH and magnetoresistance experiments show that
at low densities both $m^{\ast }$ and $\chi _{s}^{\ast } $ are
significantly enhanced compared to their band values
\cite{pudalov_review} and, according to some studies
\cite{vitkalov,shashkin1}, even diverge at the resistive
transition point.

Although none drastically non-FL features of the metallic state
have been found in ShdH measurements as of now, there is one very
intriguing observation which does seem to present a challenge for
the FL theory, at least in its conventional formulation.  Namely,
in all studies when the spin and orbital degrees of freedom were
controlled independently by applying a
tilted magnetic field, the effective masses, $m_{\uparrow }^{\ast }$ and $%
m_{\downarrow }^{\ast },$ and Dingle temperatures (impurity
scattering rates), $T_{D\uparrow }$ and $T_{D\downarrow },$ of
spin-up and -down electrons, were found to be almost the
\emph{same}. Moreover, $m^*$ in MOSFETs \cite{pud_paper, shashkin2} was found
to be independent of the spin polarization, whereas $T_D$ was
shown to depend on the polarization only weakly. In n-GaAs,
the effective mass was found to depend on the parallel magnetic field \cite{shayegan1}%
; however, this behavior was attributed to the coupling between
the in- and out-of-plane degrees of freedom (Stern effect
\cite{stern}), which is to be expected in systems with wider
quantum wells. Given that the Stern effect is subtracted off, the
resulting dependence of $m^{\ast }$ on the polarization is likely
to be weak.

Why is this strange? Polarization is expected to lead to two
effects: the spin-splitting of the effective mass, i.e.,
$m_{\uparrow }^{\ast }\neq m_{\downarrow }^{\ast }$, and
dependences of both $m_{\uparrow }^{\ast }$ and $m_{\downarrow
}^{\ast }$ on the polarization. The first effect can be understood
by considering a partially spin-polarized FL as a two-component
system. As the densities of the components are different, the
corresponding couplings describing the interactions between the
same and opposite spins are also different; hence \textit{a
priori} the mass renormalizations should also be different. That
the masses should depend on polarization can be seen from
considering two limiting cases: of zero- and full polarization. At
fixed density $n$, the Fermi energy is doubled by fully polarizing
the 2D system, hence the ratio of the Coulomb to Fermi energy
$g\equiv e^{2}\sqrt{\pi n}/E_{F}$ differs by a factor of $2$
between the cases of zero and full polarization.  The experiment
shows that the mass does depend on the density; however, if $g $
is the only dimensionless parameter that determines
the mass renormalization, the same effect can be achieved by either varying $%
n$ or by varying $E_{F}$ via polarization at fixed $n.$ Also,
different Fermi velocities should result in different impurity
scattering times for spin-up and -down electrons; hence the Dingle
temperatures are also expected to be different. However, this is
not what the experiment shows.

The qualitative arguments given above can be verified in a number
of ways. Back in 1971, Overhauser predicted the spin-splitting and
polarization dependence of $m^{\ast }$ within the RPA
approximation  for the 3D case \cite {overhauser}. Repeating the
calculation in 2D gives a similar result:
\begin{equation}
m_{\uparrow \downarrow }^{\ast }/m=1+\left( r_{s}/\sqrt{2}\pi
\right) \ln r_{s}\mp \left( r_{s}\xi /2\sqrt{2}\pi \right) \ln
r_{s}, \label{rpa_small_xi}
\end{equation}
where $\xi =(n_{\uparrow }-n_{\downarrow })/\left( n_{\uparrow
}+n_{\downarrow }\right) \ll 1$ is the polarization and $r_{s}=me^{2}/\sqrt{%
\pi n}$. In the fully-polarized regime $\left( \xi =1\right) ,$
the spin-down electrons disappear, whereas the renormalization of
$m^*_{\uparrow}$ is by a factor of $\sqrt{2}$ smaller than for
$\xi =0$. This argument can be generalized for a (partially)
spin-polarized FL \cite {meyerovich}, where the Landau interaction
function has three independent components: $f^{\uparrow \uparrow
}$, $f^{\downarrow \downarrow }$, and $f^{\uparrow \downarrow
}=f^{\downarrow \uparrow }$. The Galilean invariance then gives
\begin{eqnarray*}
m/m_{\uparrow }^{\ast } &=&1-F_{1}^{\uparrow \uparrow
}-(k_{F\downarrow
}/k_{F\uparrow })F_{1}^{\uparrow \downarrow }; \\
m/m_{\downarrow }^{\ast } &=&1-F_{1}^{\downarrow \downarrow
}-(k_{F\uparrow }/k_{F\downarrow })F_{1}^{\uparrow \downarrow },
\end{eqnarray*}
where $F_{1}^{ij}=m\int d\theta \cos \theta f^{ij}\left( \theta
\right)
/(2\pi )^{2}$, with $i,j=\uparrow ,\downarrow $. Again, in general, $%
m_{\uparrow }^{\ast }\neq m_{\downarrow }^{\ast }.$ In addition,
the spin-splitting and polarization dependence of $m^*$ are also
obtained within the Gutzwiller approximation for the Hubbard model
\cite{spalek} (in this case, mass-splitting disappears at
half-filling but the polarization dependence survives).

Absence of the polarization dependence of the effective mass suggests that $%
m^{\ast }$ is renormalized via the interaction with some classical
degree of freedom, which is not affected by the quantum degeneracy
of the electron states. In this paper, we show such a mechanism
may be provided by the interaction with (virtual) plasmons which
dominate the mass renormalization beyond the weak-coupling regime.
To this end, we turn to a model of a Coulomb gas with large
degeneracy $N$, considered previously in
Refs.~\cite{takada},~\cite {iordanski}. This model is relevant,
first of all, to valley-degenerate systems, such as the (001)
surface of a Si MOSFET, where $N=4$ (two valleys and two spin
projections). As the valley degeneracy plays a very important role
in the dirty FL theory \cite{zna, finn} it is important to
elucidate its role for the properties of a clean FL. However, the
$1/N$ expansion turns out to be converging reasonably fast even
for a non-valley degenerate system ($N=2$) and, as such, it
provides a simple yet non-trivial way of going beyond the
weak-coupling limit for not too strong Coulomb interaction.

\begin{figure}[bp]
\resizebox{.43\textwidth}{!}{\includegraphics{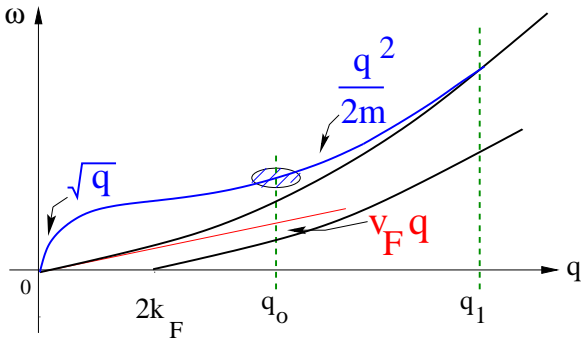}}
\vspace{0.3cm} \caption{Excitation spectrum for an $N$-fold
degenerate 2D Coulomb gas in the strong-screening regime
($r_sN^{3/2}\gg 1$). The plasmon dispersion crosses over from the
$\sqrt{q}$ to $q^2$ form at $q\sim q_0\sim r_s^{1/3}n^{1/2}\gg
k_F$. Processes with momentum and energy transfers in the shaded
oval ($q\sim q_0$ and $\omega\sim q^2_0/m$) dominate the mass
enhancement. The plasmon spectrum merges with the continuum at
$q=q_1 \sim r_s^{1/2}N^{1/4}n^{1/2}\gg q_0$. } \label{fig:bos}
\end{figure}

For a 2D $N$-fold degenerate Coulomb gas, the Fermi momentum is
scaled down by a factor of $N^{-1/2}$ (since one has to distribute
the same number of
electrons among $N$ isospin flavors), whereas the inverse screening radius ($%
\kappa $), proportional to the density of states, is scaled up by
a factor of $N.$ The ratio $\alpha \equiv \kappa
/k_{F}=r_{s}N^{3/2}/2$ controls the crossover between the regimes
of weak ($\alpha \ll 1$) and strong ($\alpha \gg 1$) screening.
For $N\gg 1,$ both of these regimes are compatible with the
condition $r_{s}\ll 1$ which guarantees that the screening cloud
includes many electrons, so that the mean-field theory is
applicable. For $\alpha\ll 1$, the screening radius
$\kappa^{-1}=\alpha^{-1} k_F^{-1}$ is larger than the Fermi
wavelength. [This case also includes the usual RPA scheme for
$N=2$--see Eq.~(\ref {rpa_small_xi}).] The mass renormalization is
mostly due to elastic scattering within the particle-hole
continuum with momentum transfers $q\sim \kappa $, whereas the
interaction with plasmons is small. In this regime, the mass
depends on total degeneracy ($N$) and is thus strongly affected by
polarization. Also, as scattering is mostly by small angles,
$m^*<m$.  For $\alpha\gg 1$, the effective screening radius
$q_0^{-1}=(2\alpha)^{-1/3} k_F^{-1}$ is {\it smaller} than the
Fermi wavelength (but still larger than the distance between
electrons); hence, scattering is isotropic ($s$-wave). The
particle-hole continuum contribution to $m^{\ast }$ is greatly
reduced for $s $-wave scattering, whereas the interaction with
virtual plasmons now plays a dominant role. As the plasmon is a
classical collective mode, it is not affected by a change in $N.$
Consequently, the leading term in the $N^{-1}$ expansion for
$m^{\ast }$ does not depend on $N, $ whereas the next-to-leading
term happens to be numerically small.

The effective mass is found from the self-energy via the usual
relation (valid for a small renormalization)
\begin{equation*}
m^{\ast }/m=1-\left( \frac{\partial \Delta \Sigma _{k}(\varepsilon )}{%
\partial \epsilon _{k}}+\frac{\partial \Delta \Sigma _{k}(\varepsilon )}{%
\partial (i\varepsilon )}\right) \Big|_{k\rightarrow k_{F},\varepsilon
\rightarrow 0},
\end{equation*}
where $\Delta \Sigma _{k}(\varepsilon )=\Sigma _{k}(\varepsilon
)-\Sigma _{k_{F}}(0)$. It is convenient to separate $\Delta \Sigma
_{k}(\varepsilon )$ into the static and dynamic parts as
\begin{equation}
\Delta \Sigma _{k}(\varepsilon )=\Delta \Sigma
_{k}^{\text{st}}(\varepsilon )+\Delta \Sigma
_{k}^{\text{dyn}}(\varepsilon ),  \label{sigma}
\end{equation}
where the static part for $\epsilon _{k}\equiv
(k^{2}-k_{F}^{2})/2m\rightarrow 0$ is
\begin{eqnarray}
\Delta \Sigma _{k}^{\text{st}}(\varepsilon ) &=&\int \frac{d\omega }{2\pi }%
\frac{d^{2}q}{(2\pi )^{2}}V_{q}(0)\left[ G_{\mathbf{k+q}}\left(
\varepsilon
+\omega \right) -G_{\mathbf{k}_{F}\mathbf{+q}}\left( \varepsilon \right) %
\right]  \notag \\
&=&\frac{m}{(2\pi) ^{2}}\epsilon _{k}\int_{0}^{2\pi }d\theta \cos
\theta V_{2k_{F}\sin \theta /2}\left( 0\right)  \label{sigma_st}
\end{eqnarray}
with $G_{\mathbf{k}}^{-1}\left( \varepsilon \right) =i\varepsilon
-\epsilon _{k}$ and
\begin{equation}
V_{q}\left( \omega \right) =\left[ q/2\pi e^{2}-\Pi _{q}\left(
\omega \right) \right] ^{-1}.  \label{v}
\end{equation}
The dynamic part is
\begin{eqnarray}
\Delta \Sigma _{k}^{\text{dyn}}(\varepsilon ) &=&\int \frac{d\omega }{2\pi }%
\frac{d^{2}q}{(2\pi )^{2}}[V_{q}(\omega )-V_{q}(0)]  \notag \\
&&\times \left[ G_{\mathbf{k+q}}\left( \varepsilon +\omega \right) -G_{%
\mathbf{k}_{F}\mathbf{+q}}\left( \varepsilon \right) \right] .
\label{sigma_dyn}
\end{eqnarray}

In what follows, we will need the following two forms of the
polarization bubble
\begin{eqnarray}
&&\Pi _{q}\left( \omega \right) =N\int \frac{d\varepsilon }{2\pi }\int \frac{%
d^{2}k}{\left( 2\pi \right) ^{2}}G_{\mathbf{k}}\left( \varepsilon \right) G_{%
\mathbf{k+q}}\left( \varepsilon +\omega \right)  \label{bubble} \\
&=&-\left\{
\begin{array}{cl}
\left( mN/2\pi \right) \left( 1-|\omega |/\sqrt{\omega ^{2}+v_{F}^{2}q^{2}}%
\right) ,\;\mathrm{for}\;q\ll k_{F}; & \notag \\
2n\varepsilon _{q}/\left( \varepsilon _{q}^{2}+\omega ^{2}\right) ,\;\mathrm{%
for}\;q\gg k_{F}, &
\end{array}
\right.
\end{eqnarray}
where $v_F=\sqrt{4\pi n/m^2N}$ and $\varepsilon _{q}\equiv
q^{2}/2m$.

In the weak-screening regime, $\Delta \Sigma
_{k}^{\text{st}}\left( \varepsilon \right) $
[Eq.~(\ref{sigma})]$m^{\ast }$ gives the main contribution to
$m^*$. To logarithmic accuracy,  $m^{\ast }/m=1+\left(
r_{s}\sqrt{N}/2\pi \right) \ln \left( r_{s}N^{3/2}\right)
+\mathcal{O}(r_{s})$ in this regime. [For $N=2$ and $\xi=0$, this
reduces back to Eq.~(\ref{rpa_small_xi})].
In this regime, the plasmon contribution to $m^{\ast }$ is a subleading, $%
\mathcal{O}(r_{s})$-term.

Now we turn to the strong-screening regime. The static screened
potential in Eq.~(\ref{sigma_st}) is evaluated for $q=2k_{F}\sin
\theta /2\leq 2k_{F}$. In this range, $V_{q}(0)=2\pi
e^{2}/(q+\kappa )$ is of the same form as in the weak-screening
regime but now $V_{q}(0)$ depends on $q$ only weakly because $q\ll
\kappa $. Consequently, the angular averaging in Eq.~(\ref
{sigma_st}) renders the static contribution to $m^*$ small:
$\left( m^{\ast }/m-1\right) ^{\text{st}}=8/3\pi N\alpha $. Using
the large-$q$ form of $\Pi$ in Eq.~(\ref{bubble}), one obtains
$V_q(0)=2\pi e^2 q^2/\left(q^3+q_0^3\right)$ for $q\gg k_F$, where
$q_0=\left(2\alpha\right)^{1/3}k_F\gg k_F$ is the inverse
screening radius in this regime \cite{iordanski}. The main
contribution to $m^{\ast }$ comes from the region of large $q$ and
$\omega $ in Eq.~(\ref{sigma_dyn}), i.e., from the plasmon region.
In the
strong-screening regime, the plasmon dispersion is given by $\omega _{p}=%
\sqrt{\varepsilon _{q}^{2}+2\pi e^{2}nq/m}$. The crossover between the $%
\sqrt{q}$ and $q^{2}$ behaviors occurs at $q\sim q_{0}$. The
plasmon runs into the continuum at $q\sim
q_{1}=k_{F}(\alpha/2)^{1/2}\gg q_{0}$. Most importantly, being the
classical collective mode, plasmon is not affected by a change in
$N$. The mass renormalization can be estimated as follows.
Typical momenta and energy transfers are of the order of  $q_{0}$ and $%
\varepsilon _{q_{0}}$, respectively; thus $V_{q_{0}}\left(
\varepsilon _{q_{0}}\right) \sim e^{2}/q_{0},$ and $G\sim \omega
^{-1}\sim\varepsilon _{q_{0}}^{-1}.$ Combining these estimates
together, one finds that $\left( m^{\ast }/m-1\right)
^{\text{dyn}}\sim \int d^{2}q\int d\omega V_{q}G^{2}\sim
r_{s}^{2/3}$, which is larger than the static contribution by
$\alpha ^{5/3}\gg 1$. To perform an actual calculation, we notice
that the plasmon contribution from the region of large $q$ to the
effective mass can be written as
\begin{equation}
m^{\ast }/m=1+\frac{i}{\pi }\int_{0}^{\infty }d\varepsilon _{q}\text{Res}%
\frac{V_{q}\left( \omega \right) }{\left( i\omega -\varepsilon
_{q}\right) ^{3}}|_{\omega =i\omega _{p}},  \label{plasmon}
\end{equation}
where only the poles of $V_{q}\left( \omega \right) $ were taken
into
account, and where we have used the expansion $\epsilon _{\mathbf{k+q}%
}=\epsilon _{k}+v_{F}q\cos \theta (1+\epsilon
_{k}/2E_{F})+\varepsilon _{q}.$
Substituting the large-$q$ form of $\Pi $ [Eq.~(\ref{bubble})] into $%
V_{q}\left( \omega \right) $ in Eq.~(\ref{plasmon}), one arrives
at the result of Ref.~\cite{iordanski} for the leading $1/N$ term
in $m^*$
\begin{equation}
m^{\ast }/m=1+Cr_{s}^{2/3},  \label{mstar_lead}
\end{equation}
where $C=\Gamma (1/3)\Gamma (1/6)/60\sqrt{\pi }\approx
0.\,\allowbreak 14$.

\begin{figure}[bp]
\resizebox{.43\textwidth}{!}{\includegraphics{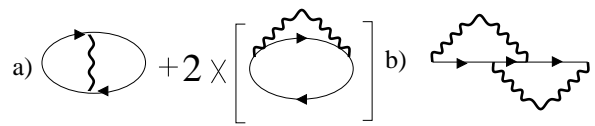}} \vspace{%
0.3cm} \caption{a: corrections to the bubble; b: vertex correction
to the self-energy.} \label{fig:2}
\end{figure}

Corrections to the leading term are obtained by including (a)
interaction corrections to the bubble [Fig.~\ref{fig:2}(a)], (b)
vertex correction to the self-energy [Fig.~\ref{fig:2}(b)], and
(c) corrections to the polarization bubble from the small-$q$
region. Estimating the diagrams in Fig.~\ref{fig:2}(a,b) in the
same way
as for the leading term, we find that both (a) and (b) contribute $N$-independent, $r_{s}^{4/3}$%
corrections to Eq.~(\ref{mstar_lead}). We have verified by an
explicit calculation that these estimates do hold. Next, we
consider correction (c) and show that it gives the next-to-leading
term in the $1/N$ expansion.

The $1/q$ correction to the large-$q$ form of the bubble
[Eq.~(\ref{bubble})] is
\begin{equation}
\delta \Pi _{q}(\omega )={}\frac{4n^{2}\pi }{mN}\frac{(3\omega
^{2}-\varepsilon _{q}^{2})\varepsilon _{q}^{2}}{(\omega
^{2}+\varepsilon _{q}^{2})^{3}}.  \label{polar_corr}
\end{equation}
At the plasmon pole ($\omega ^{2}=-\omega _{p}^{2}$) and for
$q\sim q_{0}$, the relative correction $\left| \delta \Pi
_{q}(\omega )/\Pi _{q}(\omega )\right| \sim 1/\alpha ^{2/3},$
hence one can expect the next-to-leading term in the mass to be of
order $r_{s}^{2/3}/\alpha ^{2/3}\sim 1/N.$ Indeed, a correction to
the bubble (\ref{polar_corr}) shifts the position of the
plasmon-pole from $\omega _{p}^{2}$ to $\omega _{p}^{2}+\Delta ^{2},$ where $%
\Delta ^{2}=8\pi ^{2}ne^{2}\left( 3r^{2}+1\right) /Nmqr^{4}$ and $r=\sqrt{%
1+\left( q_{0}/q\right) ^{3}}.$ Substituting this result into
Eq.~(\ref {plasmon})$,$ and evaluating the $q$-integral to
log-accuracy (the upper limit is determined by $q\sim q_{1}$,
corresponding to the region where the plasmon runs into the
continuum), we obtain $m^*$ within the next-to-leading order in
$1/N$ as
\begin{figure}[tp]
\resizebox{.33\textwidth}{!}{\includegraphics{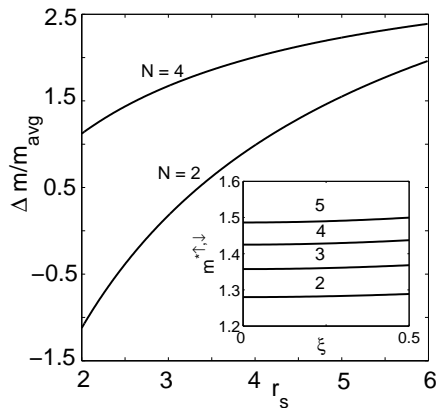}} \vspace{%
0.2cm} \caption{Change in the effective mass under full spin
polarization [cf. Eq.~(\ref{change})], as a function of $r_{s}$.
Inset: polarization dependence of the effective mass for $r_s = 2,
3, 4, 5$. } \label{fig:3}
\end{figure}
\begin{equation}
m^{\ast }/m=1+0.14r_{s}^{2/3}+\frac{1}{12N}\log \left( r_{s}N^{3/2}\right) +%
\mathcal{O}\left( \frac{1}{r_{s}N^{5/2}}\right) ,  \label{mstar}
\end{equation}
where the last term is the static contribution of the continuum. We see that the $%
1/N$ expansion generates the series in powers of
$(r_{s}N^{3/2})^{-1}$.

Now we apply our main result, Eq.~(\ref{mstar}), to real systems.
[In what follows, we neglect the last term in Eq.~(\ref{mstar}).]
First of all, due
to a small numerical coefficient in the leading term in Eq.~(\ref{mstar}%
), the actual constraint on $r_{s}$ being small is rather soft: a
two-fold enhancement of the mass occurs only for $r_{s}\approx
20,$ hence smaller values of $r_{s}$ still allows for a reasonable
description within the mean-field theory. Eq.~(\ref{mstar}) agrees
well with the observed dependence of $m^{\ast }\left( r_{s}\right)
$  for Si MOSFETs  in the range $r_s = 2-6$; for larger $r_s$, the
theoretical value of $m^{\ast }$ falls below the experimental one.
In the interval $2\leq r_{s}\leq 6$, the $1/N$ term in
Eq.~(\ref{mstar}) is not that small: it constitutes 18-26 \% and
26-32\% of the leading term for $N=4$ and $N=2,$ correspondingly.
However, the relative change in $m^{\ast }$ due to full spin
polarization ($N\rightarrow N/2$)
\begin{equation}
\frac{\Delta m}{m_{\text{avg}}}=2\times\frac{m^{\ast
}(N/2)-m^{\ast }\left( N\right) }{m^{\ast }(N/2)+m^{\ast
}(N)}\times 100\%, \label{change}\end{equation} is small. $\Delta
m/m_{\text{avg}}$ as a function of $r_s$ is shown in
Fig.~\ref{fig:3} for $N=4$  and $N=2$. In both cases, these
changes are less than 3 \%, which is likely to be below the
experimental error in the measured mass. At finite polarization,
the result in Eq.~(\ref{mstar}) changes to
\begin{equation}
\frac{m_{\uparrow ,\downarrow }^{\ast
}}{m}=1+0.14r_{s}^{2/3}+\frac{1+\xi ^{2}}{12N}\log \left[
\frac{r_{s}N^{3/2}}{1+\xi ^{2}}\right] . \label{mstar_xi}
\end{equation}
Notice that although an explicit polarization dependence does
occur in the second term, there is no spin-splitting of the masses
to this order in $1/N$. Eq.~(\ref{mstar_xi}) is valid as long as
there are still many spin-down electrons within the screening
radius or, equivalently, $1-\xi \gg r_{s}^{2/3}\sim(m^{\ast
}/m-1).$ Fig.~\ref{fig:3} shows that the effective mass remains
essentially constant in the whole range of $\xi $, which is in
agreement with the experiment~\cite {shashkin2}.

To leading order in $1/N$, the renormalization of $\chi ^{\ast }$
is entirely due to that in $m^{\ast },$ so that  $g^{\ast }=\chi
^{\ast }/m^{\ast }$ remains unrenormalized \cite{iordanski}. We
found that this remains true up to the next-to-leading term in
$1/N.$ This result is in qualitative agreement with the
experiments on Si MOSFETs. However, recent experiment on AlAs
system shows that the $g^{\ast }$ factor is affected by lifting
the valley degeneracy~\cite{shkolnikov}. More work is required to
attribute this behavior to a many-body effect.

Now, we comment briefly on the impurity scattering rate in the
large-$N$ limit. In the strong-screening regime, the screening
radius ($q_{0}^{-1}$) is much
shorter than the Fermi wavelength. Therefore, scattering even on \emph{charged%
} impurities is in the $s$-wave regime. We assume that the main
role is played by impurities within the 2D layer. Due to a
peculiarity of 2D scattering \cite{adhikari}, the scale of the
scattering cross-section section is set by the wavelength (rather
than by the impurity size $a\sim q_0^{-1}$) and depends on $a$
weakly: $A\sim k_{F}^{-1}/\ln ^{2}\left( k_{F}a\right)$.
Consequently, the scattering rate $1/\tau =n_{i}v_{F}A,$ where
$n_{i}$ is
the concentration of impurities, has only a weak dependence on the polarization (via $%
k_{F}$ under the logarithm). Thus $1/\tau$ (Dingle temperature)
for spin-up and down-electrons are close to each other. Notice
that both ShdH and weak-field Hall effect \cite{vitkalov_hall}
show that $1/\tau$, while being the same for spin-up and spin-down
electrons, increases strongly with $r_s$. Within our model, this
can only be explained by an increase in the number of scatterers
$n_i$ with decreasing electron density--not an improbable scenario
for Si MOSFETs.

Finally, we observe that as the mass is renormalized by plasmons with large $%
q$, the behavior of the plasmon spectrum at small $q$ (gapped or
gapless) is
irrelevant. Consequently, in 3D the mass is renormalized in a similar way: $%
m^*/m=1+C_{3D}r_s^{3/4}$, where $C_{3D}$ does not depend on $N$.
Therefore, a finite thickness of the 2D layer should not affect
the (approximate) spin-independence of the mass.

We aknowledge stimulating conversations with E. Abrahams, A.
Chubukov, M. Fabrizio, M. Gershenson, A. Leggett, E. Mishchenko,
V. Pudalov, M. Shayegan, B. Spivak, and S. Vitkalov. This work was supported by
NSF DMR-0308377.

\end{document}